\documentclass[11pt,a4paper]{article}
\usepackage[colorlinks=true, linkcolor=black!50!blue, urlcolor=blue, citecolor=blue, anchorcolor=blue]{hyperref}
\usepackage[font=small,labelfont=bf,margin=0mm,labelsep=period,tableposition=top]{caption}
\usepackage[a4paper,top=3cm,bottom=2.5cm,left=2.5cm,right=2.5cm,bindingoffset=0mm]{geometry}

\usepackage{graphicx,placeins}
\usepackage{float}
\usepackage{afterpage}
\usepackage{epsfig,cite}
\usepackage{amssymb}
\usepackage{amsmath}
\usepackage{dsfont}
\usepackage{multirow}
\usepackage{url}
\usepackage{xcolor,colortbl}
\usepackage{float}
\usepackage{afterpage}
\usepackage{url}
\usepackage{hyperref}
\usepackage{booktabs}
\usepackage{mathrsfs}
\usepackage{listings}

\usepackage{tikz}
\usepackage{tikz-3dplot}
\usepackage[compat=1.0.0]{tikz-feynman}
\usetikzlibrary{positioning}

\usepackage{enumitem}
\usepackage{hyperref}
\usepackage{cite}

\usepackage{pifont}

\usetikzlibrary{shapes, arrows}
\usetikzlibrary{decorations.pathreplacing}
\usetikzlibrary{positioning, calc}
\tikzstyle{fitted} = [rectangle, minimum width=5cm, minimum height=1cm, text centered, draw=black, fill=red!30]
\tikzstyle{operations} = [rectangle, rounded corners, minimum width=2cm,text centered, draw=black, fill=red!30]
\tikzstyle{roundtext} = [rectangle, rounded corners, minimum width=2cm, minimum height=0.8cm, text centered, draw=black, fill=red!30]
\tikzstyle{n3py} = [rectangle, rounded corners, minimum width=3cm, minimum height=1cm, text centered, draw=black, fill=green!30]
\tikzstyle{myarrow} = [thick,->,>=stealth]
\tikzstyle{line} =[draw, -latex']
\tikzstyle{decision} = [diamond, draw, fill=red!20, text width=7.5em, text centered,  inner sep=0pt, minimum height=2em, aspect=4]
\tikzstyle{cloud} = [draw, ellipse,fill=green!20, minimum height=2em]
\tikzstyle{inout} = [rectangle, draw, fill=green!20, text width=9.5em, text centered, rounded corners, minimum height=2em, minimum width=10em]
\tikzstyle{block}=[rectangle, draw, fill=blue!20, text width=9.5em,
                   text centered, rounded corners, minimum height=2em,
                   minimum width=10em]

\definecolor{darkgreen}{rgb}{0.0, 0.5, 0.13}

\definecolor{codegreen}{rgb}{0,0.6,0}
\definecolor{codegray}{rgb}{0.5,0.5,0.5}
\definecolor{codepurple}{rgb}{0.58,0,0.82}
\definecolor{backcolour}{rgb}{0.97,0.97,0.98}
\lstdefinestyle{mystyle}{
    backgroundcolor=\color{backcolour},
    commentstyle=\color{codegreen},
    keywordstyle=\color{magenta},
    numberstyle=\tiny\color{codegray},
    stringstyle=\color{codepurple},
    basicstyle=\ttfamily\footnotesize,
    breakatwhitespace=false,
    breaklines=true,
    captionpos=b,
    keepspaces=true,
    numbers=left,
    numbersep=5pt,
    showspaces=false,
    showstringspaces=false,
    showtabs=false,
    tabsize=2,
    frame=tb,
    framerule=0.1pt
}
\newcommand{\be}{\begin{equation}}
\newcommand{\ee}{\end{equation}}
\newcommand{\bea}{\begin{eqnarray}}
\newcommand{\eea}{\end{eqnarray}}
\newcommand{\bi}{\begin{itemize}}
\newcommand{\ei}{\end{itemize}}
\newcommand{\ben}{\begin{enumerate}}
\newcommand{\een}{\end{enumerate}}

\def\gsim{\mathrel{\rlap{\lower4pt\hbox{\hskip1pt$\sim$}}
    \raise1pt\hbox{$>$}}}         %
\def\lsim{\mathrel{\rlap{\lower4pt\hbox{\hskip1pt$\sim$}}
    \raise1pt\hbox{$<$}}}         %

\newcommand{\draft}[1]{}

\def\beq{\begin{equation}}
\def\eeq{\end{equation}}

\def\lapprox{\lower .7ex\hbox{$\;\stackrel{\textstyle <}{\sim}\;$}}
\def\gapprox{\lower .7ex\hbox{$\;\stackrel{\textstyle >}{\sim}\;$}}

\def\GeV{{\rm GeV}}

\def\d{{\rm d}}

\numberwithin{equation}{section}
\numberwithin{figure}{section}
\numberwithin{table}{section}

\usepackage{tabularx}
\newcolumntype{C}[1]{>{\centering\arraybackslash}p{#1}}

\RequirePackage[status=draft,inline,nomargin,marginclue]{fixme}
\fxusetheme{color}
\FXRegisterAuthor{fh}{afh}{FH}
\FXRegisterAuthor{ac}{aac}{AC}
\FXRegisterAuthor{ab}{aab}{AB}
\FXRegisterAuthor{gm}{agm}{GM}
\FXRegisterAuthor{rs}{ars}{RS}

\usepackage{amsmath}
\usepackage{amsfonts}
\usepackage{amssymb}
\usepackage{dsfont}
\usepackage{pifont}
\usepackage{booktabs}
\usepackage{graphicx}
\usepackage{epstopdf}
\usepackage{epsfig}
\usepackage{framed}
\usepackage{makeidx}
\usepackage{siunitx}
\usepackage[capitalise]{cleveref}
\usepackage{hyperref}
\usepackage{placeins}
\usepackage[font=small,labelfont=bf]{caption}

\RequirePackage{csquotes}
\usepackage{bbm}

\begin{document}
\newgeometry{top=1.5cm,bottom=1.5cm,left=1.5cm,right=1.5cm,bindingoffset=0mm}

\vspace{-2.0cm}
\begin{flushright}
Nikhef-2024-014\\
Edinburgh 2024/5\\
TIF-UNIMI-2024-13\\
\end{flushright}
\vspace{0.3cm}

\begin{center}
  {\Large \bf An FONLL prescription with coexisting flavor number PDFs}
  \vspace{1.1cm}

  Andrea Barontini$^1$,
  Alessandro Candido$^{1,2}$,
  Felix Hekhorn$^{1,3,4}$,
  Giacomo Magni$^{5,6}$,
  and Roy Stegeman$^7$

  \vspace{0.7cm}

  {\it \small
    ~$^1$Tif Lab, Dipartimento di Fisica, Universit\`a di Milano and\\
    INFN, Sezione di Milano, Via Celoria 16, I-20133 Milano, Italy\\[0.1cm]
    ~$^2$CERN, Theoretical Physics Department, CH-1211 Geneva 23, Switzerland\\[0.1cm]
    ~$^3$University of Jyvaskyla, Department of Physics, P.O. Box 35, FI-40014 University of Jyvaskyla, Finland\\[0.1cm]
    ~$^4$Helsinki Institute of Physics, P.O. Box 64, FI-00014 University of Helsinki, Finland\\[0.1cm]
    ~$^5$Department of Physics and Astronomy, Vrije Universiteit, NL-1081 HV Amsterdam\\[0.1cm]
    ~$^6$Nikhef Theory Group, Science Park 105, 1098 XG Amsterdam, The Netherlands\\[0.1cm]
    ~$^7$The Higgs Centre for Theoretical Physics, University of Edinburgh,\\
    JCMB, KB, Mayfield Rd, Edinburgh EH9 3JZ, Scotland\\[0.1cm]
  }

  \vspace{0.7cm}

  {\bf \large Abstract}

\end{center}

We present a new prescription to account for heavy quark mass effects in the determination
of parton distribution functions (PDFs) based on the FONLL scheme.
Our prescription makes explicit use of the freedom to choose the number of active flavors at a given scale
and, thus, use coexisting PDFs with different active flavor number.
This new prescription is perturbatively equivalent to the former but improves the implementation in two ways.
First, it can be naturally generalized to account simultaneously for multiple heavy quark effects, such as charm and bottom effects,
which can both be relevant at the same scale due to the small mass difference.
Second, it can be trivially generalized to use at any fixed-order or collinear resummed accuracy,
while previous prescriptions required ad-hoc expansions of the DGLAP evolution kernels for each coefficient.
We supplement the paper with codes for the computation of deep inelastic scattering observables
in this new prescription.
\clearpage

\tableofcontents

\section{Introduction}
\label{sec:intro}

The high energy physics community has seen significant progress in increasing both the
theoretical and experimental precision in the last few years~\cite{Amoroso:2022eow}.
In the context of hadronic scattering, one of the major source of uncertainties
in the prediction of accurate observables comes from the determination of parton distribution functions (PDFs).
Recently, several possible lines of improvement have been investigated, from the inclusion of
photon evolution effects~\cite{Cridge:2021pxm,NNPDF:2024djq}, to the extension to
next-to-next-to-next-to-leading order (N$^3$LO) in the perturbative strong
coupling~\cite{Caola:2022ayt,McGowan:2022nag,Cridge:2023ryv,NNPDF:2024nan,Barontini:2024dyb,Cooper-Sarkar:2024crx}
and the inclusion of theoretical uncertainties~\cite{NNPDF:2019ubu,NNPDF:2024dpb}.
In this work, we are interested in the description of
heavy quark mass effects, especially in deep inelastic scattering (DIS),
which can become significant if the relevant process scale is of the order of any heavy quark mass.

For instance, in the case of the HERA data these corrections turned out to be up to $\SI{20}{\percent}$
of the observed cross-section \cite{H1:2015ubc,H1:2018flt} and this feature is also expected at the planned
Electron-Ion Collider in the US~\cite{AbdulKhalek:2021gbh} and China~\cite{Anderle:2021wcy}, as both
are designed as low scale colliders. However, also for experiments at the LHC, these corrections
can be of great relevance if the scale of the process of interest is low, e.g., B-meson production~\cite{Helenius:2023wkn,Cacciari:2012ny}.

As there is no unique prescription on how to include heavy quark mass effects into the theoretical
calculations, several approaches have been suggested~\cite{Aivazis:1993pi,Thorne:1997uu,Kramer:2000hn,Tung:2001mv,Nadolsky:2009ge,Forte:2010ta,Guzzi:2011ew}.
The general idea of these schemes, known as general mass variable flavor number schemes (GM-VFNSs),
is to combine fixed order calculations, which retain all heavy quark mass effects,
and collinear resummed calculations.
The usage of GM-VFNSs is necessary due to the finite perturbative knowledge and as such all schemes attempt to mimic the
exact analytic behavior.
Here, we focus on the FONLL approach which was originally suggested for heavy flavor hadroproduction~\cite{Cacciari:1998it},
later applied to DIS structure functions at NLO~\cite{Forte:2010ta} and extended to intrinsic charm~\cite{Ball:2015dpa}.

We propose a new prescription of the FONLL scheme which is perturbatively equivalent to the former,
but is capable of dealing with an arbitrary accuracy in either the fixed-order calculations
or the collinear resummed calculations and also allows a direct
application to the case of any number of parton distributions.

Indeed, a complication with the original prescription~\cite{Forte:2010ta} is the way in which the final
coefficient functions are constructed: the procedure rewrites all expression using a single PDF
(as is the case for all heavy quark mass schemes so far), which results in a prescription that is non-trivial
to follow in practice. This problem becomes apparent when dealing with high perturbative orders (i.e.\ N$^3$LO)
or with hadronic collision, e.g.\ at the LHC, with multiple PDFs involved.
The assumptions of using only a single PDF is however not required in FONLL and our new approach primarily relies on
lifting this assumption.

Furthermore, the former FONLL prescription only considers a single mass problem without giving a clear recipe on how to deal
with the multi-mass case. In practice this is, however, a relevant question as the charm quark mass and the
bottom quark mass are of similar order of magnitude. We address this issue specifically and show
how our new approach can resolve the issue in a natural way.

By using the same initial PDF defined in different flavour number schemes, instead of rewriting the expressions in terms of a single PDF, we achieve a clear separation between evolution and partonic matrix elements.
The \texttt{EKO} package~\cite{Candido:2022tld} for solving the DGLAP evolution
equations, supports the computation of such coexisting flavor number PDFs for a given factorization scale, while the \texttt{yadism}~\cite{Candido:2024rkr} library allows for the computation of DIS structure functions.
Together these codes allow for the evaluation of DIS structure functions in this new FONLL prescription.
Coexisting PDFs computed with \texttt{EKO} could be used to compute observables for other processes in the FONLL scheme, but we do not provide such codes.

The rest of the paper proceeds as follows.
In \cref{sec:theory} we establish the notation used in this paper and recall the necessary ingredients to perform
PDF evolution. In \cref{sec:one} we explicitly construct the FONLL scheme
for DIS structure functions for the case of a single heavy quark following the procedures of Ref.~\cite{Forte:2010ta} and the new way proposed in this paper.
In \cref{sec:two} we examine the case of multiple heavy quarks and in \cref{sec:damping}
we combine the two cases to a prescription valid at all scales.
In \cref{sec:schemes} we discuss the generalization to arbitrary accuracy of the perturbative
calculation or of the resummed calculation, to intrinsic heavy quark treatment~\cite{Ball:2022qks} and to arbitrary observables.
In \cref{sec:summary} we summarize our results and, finally, we provide two appendices with further details.
In \cref{app:b} we comment on a generalized treatment of the two mass case and
in \cref{app:implementation} we briefly comment on the practical
implementation of our new prescription in the \texttt{pineline} framework~\cite{Barontini:2023vmr}.

\section{Constructing Flavor Number Schemes}
\label{sec:theory}

High energy hadronic scattering can be described using the collinear factorization theorem~\cite{Collins:1989gx}.
In doing so, the physical observable is computed as a convolution
between the short-distance (high scale) contribution, given by the
partonic matrix elements, and a long-distance (low scale)
contribution, given by the universal PDFs.
The former is computed at a given accuracy using perturbative QCD,
while the latter is intrinsically related to how the partons
are distributed inside the colliding hadron.
The dependence of PDFs $f(x,\mu^2)$ on the factorization scale $\mu^2$,
is determined by the associated renormalization group equation (RGE),
commonly referred to as DGLAP equations~\cite{ap,dok,gl},
whose kernels admit a perturbative QCD expansion.

When solving the DGLAP equations
or computing the partonic matrix elements,
we need to consider the necessary Feynman diagrams and
also make an assumption on which flavors
may, or may not, participate in each line of the diagram.
The value of the active number of flavors is typically associated
to the relevant scale $Q^2$ of the process, but there is no unique
prescription on how to treat the heavy quark flavors.
A specific resolution of this ambiguity is referred to as a flavor number scheme (FNS).
In particular, to define a FNS it is sufficient consider the masses of the quarks
which can take three different states: light ($m=0$), heavy ($m$ finite), and decoupled ($m = \infty$).
The option of using a configuration in which the number of contributing flavors is constant, i.e.\ it
does not depend on any scale, is called the fixed flavor number scheme (FFNS).
We denote a fixed flavor number scheme with just $n_f$ light flavors by FFNS$n_f$.
If a FFNS also accounts for finite mass contributions of the heavy quarks we add the corresponding symbols as suffixes (FFNS$n_fh$).
Our notation differs from that found in previous literature where the heavy quark contributions are usually not explicitly denoted in the symbol for the FNS, as such we give some examples of our notation in \cref{tab:ffns}.

\begin{table}
    \centering
    \begin{tabularx}{.68\linewidth}{l|c|c|c|c}
        \toprule
        Scheme name & up/down/strange & charm & bottom & top\\
        \midrule
        FFNS3 & light & decoupled & decoupled & decoupled\\
        FFNS3c & light & massive & decoupled & decoupled\\
        FFNS3cb & light & massive & massive & decoupled\\
        FFNS4 & light & light & decoupled & decoupled\\
        FFNS4b & light & light & massive & decoupled\\
        FFNS5 & light & light & light & decoupled\\
        \bottomrule
    \end{tabularx}
    \caption{Mapping of several example configuration of a FFNS to the respective quark masses.
        Quarks can be either light ($m=0$), massive ($m$ finite), or decoupled ($m=\infty$).}
    \label{tab:ffns}
\end{table}

Let us now analyze, in more detail, how the strong coupling and PDFs are
evolved in different FNSs.
In the following, when using FFNS$n_f$, we add an explicit superscript $(n_f)$
to all ingredients, which are computed with this configuration.
First, we consider the RGE of the strong coupling $a_s(\mu^2)$, the $\beta$ function,
\begin{equation}
    \mu^2\frac{\d a_s^{(n_f)}(\mu^2)}{\d \mu^2} = \beta^{(n_f)}\left(a_s^{(n_f)}(\mu^2)\right) = - \sum_{k=0}^{\infty} \left(a_s^{(n_f)}(\mu^2)\right)^{2+k} \beta^{(k),(n_f)} \ ,
    \label{eq:betafnc}
\end{equation}
where the coefficients of the beta function $\beta^{(k),(n_f)}$ are known up to
5-loop~\cite{Baikov:2016tgj,Chetyrkin:2017bjc,Herzog:2017ohr,Luthe:2017ttg}.
For any arbitrary final scale $\mu^2$, we can solve \cref{eq:betafnc}
with a given boundary condition $a_s^{(n_f)}(\mu_0^2)$, at the initial scale $\mu_0^2$,
in terms of a couplings operator $T$
\begin{equation}
    a_s^{(n_f)}(\mu^2) = T^{(n_f)}(\mu^2 \leftarrow \mu_0^2) a_s^{(n_f)}(\mu_0^2) \,.
    \label{eq:cko}
\end{equation}

In an analogous way the DGLAP equations, are given by
\begin{equation}
    \mu^2\frac{\d f_i^{(n_f)}(\mu^2)}{\d \mu^2} = -\gamma_{ij}^{(n_f)}\left(a_s^{(n_f)}(\mu^2)\right) f_j^{(n_f)}(\mu^2) = - \sum_{k=0}^{\infty} \left(a_s^{(n_f)}(\mu^2)\right)^{1+k} \gamma_{ij}^{(k),(n_f)} f_j^{(n_f)}(\mu^2) \,,
    \label{eq:DGLAP}
\end{equation}
where $i,j$ are the flavor indices, here and in the rest of the paper, the standard convolution in momentum space
is not written explicitly.
The coefficients of the anomalous dimension $\gamma_{ij}^{(k),(n_f)}$ are known up to approximate 4-loop order~\cite{Davies:2016jie,Moch:2017uml,
Davies:2022ofz,Henn:2019swt,Bonvini:2018xvt,Moch:2021qrk,Soar:2009yh,McGowan:2022nag,Falcioni:2023luc,Hekhorn:2023gul,NNPDF:2024nan}.
Again, we can solve the RGEs with a given boundary value $f_i^{(n_f)}(\mu_0^2)$ with an evolution kernel operator (EKO) $E$ via
\begin{equation}
    f_i^{(n_f)}(\mu^2) = E_{ij}^{(n_f)}(\mu^2 \leftarrow \mu_0^2) f_j^{(n_f)}(\mu_0^2) \,.
    \label{eq:eko}
\end{equation}

\cref{eq:betafnc,eq:DGLAP} arise from the cancellation of ultra-violet (UV)
or collinear poles associated to the $n_f$ light quarks, respectively. This leads to the logarithmic nature of the RGEs.
Indeed, both RGEs correspond to a resummation:
due to the logarithmic differential on the l.h.s., when solving the equations,
we collect logarithms of the form $\log(\mu^2/\mu_0^2)$.
Other than these logarithms there is no dependence on the scale, i.e.\
both the coupling operator $T$ and the EKO $E$ are only functions
of $\log(\mu^2/\mu_0^2)$.
We stress that to obtain a FFNS$n_f$ result, we need to use a consistent configuration
throughout, i.e.\ the boundary conditions are given in FFNS$n_f$ and the operators use FFNS$n_f$ ingredients.

Next, we move to the computation of observables: to simplify the discussion in the beginning we consider an observable $F$
which depends linearly on a PDF $f$. In \cref{sec:schemes} we explicitly remove this assumption and show how to construct the new prescription
for any observable. In an FFNS setup, the observable $F$ can be computed as a convolution between the respective PDF
$f^{(n_f)}$ and a coefficient function $C^{(n_f)}$,
\begin{equation}
    F^{(n_f)} = f^{(n_f)} C^{(n_f)}\quad\text{with}~C^{(n_f)}(x) = \sum_k \left(a_s^{(n_f)}\right)^k C^{(n_f)}_k(x) \, ,
\end{equation}
where we expanded the coefficient function into a power series of $a_s^{(n_f)}$. Note that we suppressed any flavor dependence
on the PDF or the coefficient function for the sake of clarity and that the lower bound of the series expansion depends on the
observable $F$. When computing the higher order contributions to the coefficient functions $C^{(n_f)}_k$ one always encounters
collinear configuration which have to be reabsorbed into the PDF definition and using the $\overline{\rm MS}$ scheme
this can leave logarithms of $\ln(1-x)$ and $\ln(x)$ in the coefficient functions due to over-subtractions~\cite{Candido:2020yat}.

The simplest way to include heavy quark mass effects is then to move to a FFNS$n_fh$ scheme and to thus add an explicit mass dependency
to the coefficient functions. The coefficient functions contain again collinear configuration, which, however, do not manifest as poles when they
involve heavy quarks but instead yield collinear logarithms of the form $\ln(Q^2/m_h^2)$ inside the coefficient functions $C^{(n_fh)}_k$.
Eventually, due to the finite perturbative expansion in $a_s$, we only get a finite number of
such logarithms which turn the calculation unstable in the region of large-$Q^2$.
By allowing one light flavor more in the RGEs, in the region where $Q^2 \gg m_h^2$, these logarithms are properly resummed.
A variable flavor number scheme (VFNS) is a FNS where the number of active flavours participating in the evolution equations is dynamic.

It is thus important to use a FNS which consistently combines several FFNSs at a given matching scale.
A common choice of matching scales are the heavy quark masses such that logarithms of the form $\ln(Q^2/m_h^2)$ are included in the
region where $Q>\lambda m_h$ for some constant $\lambda$.
When transitioning from one set of active flavors to the next,
we need to apply a matching operator which can be computed perturbatively.
For matching the strong coupling $a_s$ at $\lambda_\alpha m_h^2$ we find
\begin{align}
    \begin{split}
    a_s^{(n_f+1)}(\lambda_\alpha m_h^2) &= \zeta^{(n_f+1)}\left(a_s^{(n_f+1)}(\lambda_\alpha m_h^2), \ln(\lambda_\alpha) \right) a_s^{(n_f)}(\lambda_\alpha m_h^2)  \\
        &=\sum_{k=0}^{\infty} \left(a_s^{(n_f+1)}(\lambda_\alpha m_h^2)\right)^k \sum_{l=0}^k \zeta^{kl,(n_f+1)} \ln^k(\lambda_\alpha) a_s^{(n_f)}(\lambda_\alpha m_h^2),
   \label{eq:asmatching}
    \end{split}
\end{align}
where the decoupling constants $\zeta^{kl,(n_f)}$ are known up to 4-loop~\cite{Chetyrkin:2005ia,Schroder:2005hy},
while for matching the PDFs at $\lambda_f m_h^2$ we find
\begin{align}
    \begin{split}
    f_i^{(n_f + 1)}(\lambda_f m_h^2) &= A_{ij}^{(n_f+1)}\left(a_s^{(n_f+1)}(\lambda_f m_h^2), \ln(\lambda_f)\right) f_i^{(n_f)}(\lambda_f m_h^2) \\
     &= \sum_{k=0}^{\infty} \left(a_s^{(n_f+1)}(\lambda_f m_h^2)\right)^k \sum_{l=0}^k A_{ij}^{kl,(n_f+1)}\ln^l(\lambda_f) f_j^{(n_f)}(\lambda_f m_h^2)
     \label{eq:pdfmatching}
    \end{split}
\end{align}
where the coefficients of the matching matrix $A_{ij}^{kl,(n_f)}$ are known up to 4-loop order~\cite{pdfnnlo,
Bierenbaum:2008yu,Bierenbaum:2009zt,Bierenbaum:2009mv,Ablinger:2010ty,Ablinger:2014vwa,Ablinger:2014uka,Behring:2014eya,
Ablinger_2014,Ablinger:2014nga,Ablinger_2015,Blumlein:2017wxd,Ablinger:2022wbb,Ablinger:2023ahe,Ablinger:2024xtt}.
In the following we assume that the matching is performed at the mass, $\lambda_\alpha = 1 = \lambda_f$,
but we comment on the general case, which can be easily recovered in our new prescription, in \cref{sec:schemes}.

\definecolor{mpl-blue}{RGB}{31, 119, 180}
\definecolor{mpl-orange}{RGB}{255, 127, 14}
\definecolor{mpl-green}{RGB}{44, 160, 44}
\definecolor{mpl-red}{RGB}{214, 39, 40}

\begin{figure}
  \centering
  \begin{tikzpicture}[xscale=2.5, yscale=1.5]

    \draw[line width=3.5pt] (1, 1) -- (2, 1);
    \draw[line width=3.5pt, opacity=0.4] (2, 1) -- (5, 1);
    \draw[opacity=0.4, -stealth, line width=3.5pt] (5, 1) -- (5.3, 1);

    \draw[line width=3.5pt, opacity=0.4] (1, 2) -- (3, 2);
    \draw[line width=3.5pt] (2, 2) -- (3, 2);
    \draw[line width=3.5pt, opacity=0.4] (3, 2) -- (5, 2);
    \draw[opacity=0.4, -stealth, line width=3.5pt] (5, 2) -- (5.3, 2);

    \draw[line width=3.5pt, opacity=0.4] (1, 3) -- (3, 3);
    \draw[line width=3.5pt] (3, 3) -- (4, 3);
    \draw[line width=3.5pt, opacity=0.4] (4, 3) -- (5, 3);
    \draw[opacity=0.4, -stealth, line width=3.5pt] (5, 3) -- (5.3, 3);

    \draw[line width=3.5pt, opacity=0.4] (1, 4) -- (4, 4);
    \draw[line width=3.5pt] (4, 4) -- (5, 4);
    \draw[-stealth, line width=3.5pt] (5, 4) -- (5.3, 4);

    \draw[dotted, line width=1pt] (4, 1) -- (4, 4);
    \draw[dotted, line width=1pt] (3, 1) -- (3, 4);
    \draw[dotted, line width=1pt] (2, 1) -- (2, 4);

    \node at (4, 4.5) {$m_t$};
    \node at (3, 4.5) {$m_b$};
    \node at (2, 4.5) {$m_c$};

    \node at (0.4, 1) {$n_l=3\,, n_h=1$};
    \node at (0.4, 2) {$n_l=4\,, n_h=0$};
    \node at (0.4, 3) {$n_l=5\,, n_h=0$};
    \node at (0.4, 4) {$n_l=6\,, n_h=0$};

    \draw[->] (4.4, .5) -- (5, .5) node[midway, below] {$Q$};
  \end{tikzpicture}
  \caption{}
  \caption{Schematic representation of a VFNS. 
  Each horizontal line indicates a FFNS scheme with $n_l$ light flavor and $n_h$ massive quarks. 
  Vertical lines can coexist for any scale Q denoted on the abscissa axis; 
  the solid black line corresponds to a VFNS, which changes FNS at a scale 
  equal to the quark threshold, see \cref{eq:pdfmatching}: for scales below $Q \leq m_c$ it accounts for 
  massive charm corrections, while otherwise treats all the quarks a massless, with 
  4, 5 or 6 active flavors.}
  \label{fig:fnslines}
\end{figure}

The central observation of this paper is that, at any scale $Q$, we can theoretically use any FFNS
and, moreover, we could even use them \textit{simultaneously}.
To visualize this concept, we provide \cref{fig:fnslines}
where we indicate the different FFNSs by the horizontal lines.
The RGEs, \cref{eq:cko,eq:eko}, allow to navigate in the scale $Q$, i.e.\ along the horizontal axis,
while the matching equations, \cref{eq:asmatching,eq:pdfmatching},
allow to navigate in the number of flavors $n_f$, i.e.\ along the vertical axis.
So far most evolution codes do not consider this coexistence of FNS
and hence provide no native interface to access them, but our recently developed DGLAP solver
EKO~\cite{Candido:2022tld} explicitly allows for this possibility.

Before moving to the explicit FONLL prescription, we list another VFNS example, the simple but common choice
of the zero mass-variable flavor number scheme (ZM-VFNS).
In the ZM-VFNS quarks are treated as decoupled below their respective matching scale and
light above. We can thus write for an arbitrary observable $\sigma(Q^2)$ at a given scale $Q$
\begin{equation}
    \sigma^{\text{ZM-VFNS}}(Q^2) = \left\{ \begin{array}{ll}
        \sigma^\mathrm{(3)}(Q^2) &\text{if}~ Q^2 < m_c^2\\
        \sigma^\mathrm{(4)}(Q^2) &\text{if}~ m_c^2 \leq Q^2 < m_b^2\\
        \sigma^\mathrm{(5)}(Q^2) &\text{if}~ m_b^2 \leq Q^2 < m_t^2\\
        \sigma^\mathrm{(6)}(Q^2) &\text{if}~ Q^2 \geq m_t^2
    \end{array} \right. \label{eq:zmvfns}
\end{equation}
This scheme usually generates unphysical discontinuities where the scale $Q$
equals any of the heavy quark masses. It does not account for mass effects of heavy quarks in coefficient functions,
which simplifies the calculation of the observable,
however these effects can be significant around $Q^2\sim m_h^2$.

\section{FONLL with coexisting flavor number PDFs}
Here we discuss the FONLL prescription applied to DIS structure functions. First, in \cref{sec:one} we review the prescription accounting for the massive contributions of a single heavy quark as introduced in Refs.~\cite{Forte:2010ta,Ball:2015dpa}. Then, in \cref{sec:two} we discuss the case of accounting for the massive contributions of two heavy quarks.
As in the previous section, we assume that the matching scale is equal to the quark masses, $\lambda_f = 1$, though the general case is easily recovered.
While the FONLL prescription may be applied to other observables, as discussed in \cref{sec:schemes}, in this section we focus the discussion on DIS.

\subsection{Single mass case}
\label{sec:one}

In the following we consider, without loss of generality, the case of the charm quark and assume
that the boundary conditions for the strong coupling $a_s$ and the PDF $f$ are given in the three flavor
scheme at the charm mass, i.e.\ $a_s^{(3)}(m_c)$ and $f^{(3)}(m_c)$.

The main idea of the FONLL prescription is to enhance the fixed order calculation by accounting for the resummation of collinear logarithms, which can become arbitrarily large.
In practice we sum the observables defined in FFNS3c and FFNS4, while taking care of the double counting and write
\begin{equation}
    F^\mathrm{FONLL}(Q^2,m_c^2) = F^{(3c)}(Q^2,m_c^2) + F^{(4)}(Q^2) - F^{(3c\cap 4)}(Q^2,m_c^2) \, ,
    \label{eq:FONLL1}
\end{equation}
where we suppress the $x$ dependence and borrow the intersection operation
$\cap$ from set theory language to indicate the subset of terms present in both
schemes\footnote{ In the notation of Refs.~\cite{Forte:2010ta,Ball:2015dpa} the term
$F^{(3c\cap 4)}$ is denoted by $F^{(3,0)}$.}. Each ingredient obeys factorization
and is thus given by a convolution between a PDF $f$ and a coefficient function
$C$.

In FFNS3c, the observable is given by
\begin{equation}
    F^{(3c)} (Q^2, m_c^2) = C^{(3c)}(Q^2/m_c^2) f^{(3c)}(Q^2) = C^{(3c)}(Q^2/m_c^2) E^{(3c)}(\log(Q^2/m_c^2)) f^{(3)}(m_c^2) \label{eq:sF3}
\end{equation}
with the coefficient function $C^{(3c)}$, which can depend both power-like and logarithmically on the ratio $Q^2/m_c^2$,
and the EKO $E^{(3c)}$, which contributes only logarithmically. Note that both, the EKO $E^{(3c)}$
and the coefficient function $C^{(3c)}$ also depend on the strong coupling $a_s^{(3c)}$ defined in FFNS3c, but for the sake
for readability we suppress this dependency.

Analogously, in FFNS4 the observable is given by
\begin{equation}
    F^{(4)} \left(Q^2\right) = C^{(4)} f^{(4)}\left(Q^2\right) = C^{(4)} E^{(4)}\left(\log\left(Q^2/m_c^2\right)\right) A^{(4)}\left(\log\left(\mu_c^2/m_c^2\right)=0\right) f^{(3)}\left(m_c^2\right) \label{eq:sF4}
\end{equation}
with the coefficient function $C^{(4)}$, which is independent of the charm mass as charm is considered
light, and the EKO $E^{(4)}$, which encodes the logarithmic dependency including that on the charm mass.
We abbreviate the dependency of the matching condition $A^{(4)}$ to the logarithmic ratio between the matching scale
$\mu_c^2 = \lambda_f m_c^2$ and the heavy quark mass $m_c^2$.
Remember, that here we assumed the matching scale to be at the corresponding quark mass,
thus $\lambda_f=1$, and $A^{(4)}(0)$ indicates the non-trivial scale independent part of the matching between FFNS4 and FFNS3c.
Again, we suppress the dependency of all elements on the strong coupling $a_s^{(4)}$ for readability.

Finally, we can determine the overlap, $F^{(3c\cap 4)}$, between the two expressions, \cref{eq:sF3,eq:sF4}, and we find
\begin{equation}
    F^{(3c \cap 4)} (Q^2, m_c^2) = C^{(3c \cap 4)}(\log(Q^2/m_c^2)) f^{(3c)}(Q^2) = C^{(3c \cap 4)}(\log(Q^2/m_c^2)) E^{(3c)}(\log(Q^2/m_c^2)) f^{(3)}(m_c^2)\, ,
    \label{eq:sF34}
\end{equation}
where the coefficient function $C^{(3c \cap 4)}$ depends only logarithmically on the ratio $Q^2/m_c^2$, since \cref{eq:sF4}
has only a logarithmic dependency.
Moreover, we can give an explicit expression for $C^{(3c \cap 4)}$:
\begin{equation}
    C^{(3c \cap 4)}(\log(Q^2/m_c^2)) = C^{(4)} A^{(4)}(\log(Q^2/m_c^2))
    \label{eq:C3c4}
\end{equation}
which can be obtained from \cref{eq:sF4} by displacing the matching from $\mu_c^2 = m_c^2$ (thus $A^{(4)}(0)$) to $\mu_c^2 = Q^2$
(thus $A^{(4)}(\log(Q^2/m_c^2))$). Displacing the matching reshuffles collinear logarithms from
the resummation in $E^{(4)}$ to the matching operator $A^{(4)}$.
\cref{eq:C3c4} gives a definition of $C^{(3c \cap 4)}$ in terms of FFNS4 ingredients, i.e.\ specifically of $a_s^{(4)}$, which would
naively make \cref{eq:C3c4} use multiple schemes simultaneously (comparing l.h.s. and r.h.s.). However, we can also give
a definition based on FFNS3c by observing that $C^{(3c \cap 4)}$ may only depend logarithmically on $Q^2/m_c^2$.
Thus we can extract $C^{(3c \cap 4)}$ from $C^{(3c)}$ by only retaining the collinear logarithms $\log(Q^2/m^2)$
and neglecting the power-like contributions, i.e.\ we can write:
\begin{equation}
    C^{(3c)}(Q^2/m_c^2) = C^{(3c\cap 4)}(\log(Q^2/m_c^2)) + \tilde{C}^{(3c)} \left(m_c^2/Q^2\right)\,,
\end{equation}
where $\tilde{C}^{(3c)} \left(m_c^2/Q^2\right)$ contains only terms proportional to mass power corrections.
Note that the constant term $\mathcal O(1)$ is considered part of $C^{(3c \cap 4)}$.

The prescription of Refs.~\cite{Forte:2010ta,Ball:2015dpa} now proceeds in rewriting \cref{eq:FONLL1} in terms of a single PDF
$f^{(4)}(Q^2)$, i.e.\ by transforming the coefficient function into FFNS4:
\begin{align}
    F^\mathrm{FONLL}(Q^2,m_c^2) &= C^\mathrm{FONLL}(Q^2,m_c^2) f^{(4)}(Q^2)\\
     &= \left( C^{(4)} + B^{[3c]}(Q^2,m_c^2) - B^{[3c\cap 4]}(\log(Q^2/m_c^2)) \right) f^{(4)}(Q^2) \ ,
     \label{eq:old}
\end{align}
where $B^{[3c]}$ and $B^{[3c\cap 4]}$ are related to the FFNS3c contribution, \cref{eq:sF3}, and the overlap contribution, \cref{eq:sF34},
respectively, written as a function of FFNS4 quantities. Both can be computed by explicitly reverting the PDF scheme change, $\left(A^{(4)}(Q^2/m_c^2)\right)^{-1}$, and by transforming the strong coupling $a_s$
into the required FFNS4 scheme:
\begin{equation}
    B^{[3c]}(Q^2,m_c^2) = C^{(3c)}(Q^2,m_c^2) \left(A^{(4)}(Q^2/m_c^2)\right)^{-1}.
    \label{eq:Bdef}
\end{equation}

However, using \texttt{EKO} we can compute $f^{(3c)}(Q^2)$ and $f^{(4)}(Q^2)$ at any scale $Q^2$, which allows us to simply use
\cref{eq:sF3,eq:sF4,eq:sF34} as they are without re-expressing any terms as a function of FFNS4 quantities.
This, which is the central argument of the paper, means we are dealing
with \textit{coexising flavor number PDFs}, as we have explicitly decoupled the scale and the FNS here.
Differences between this prescription, and that of Refs.~\cite{Forte:2010ta,Ball:2015dpa} are
higher order with respect to the required perturbative accuracy.

While in the prescription of Refs.~\cite{Forte:2010ta,Ball:2015dpa} the necessary cancellations
in the low or high scale region respectively happen analytically at the
level of coefficient functions, i.e.\ \textit{before} multiplying with the PDF $f^{(4)}(Q^2)$ in \cref{eq:old},
in our new prescription this happens numerically at the level of structure functions, i.e.\ on the r.h.s. of \cref{eq:FONLL1}.

\begin{figure}
    \centering
    \includegraphics[width=\textwidth]{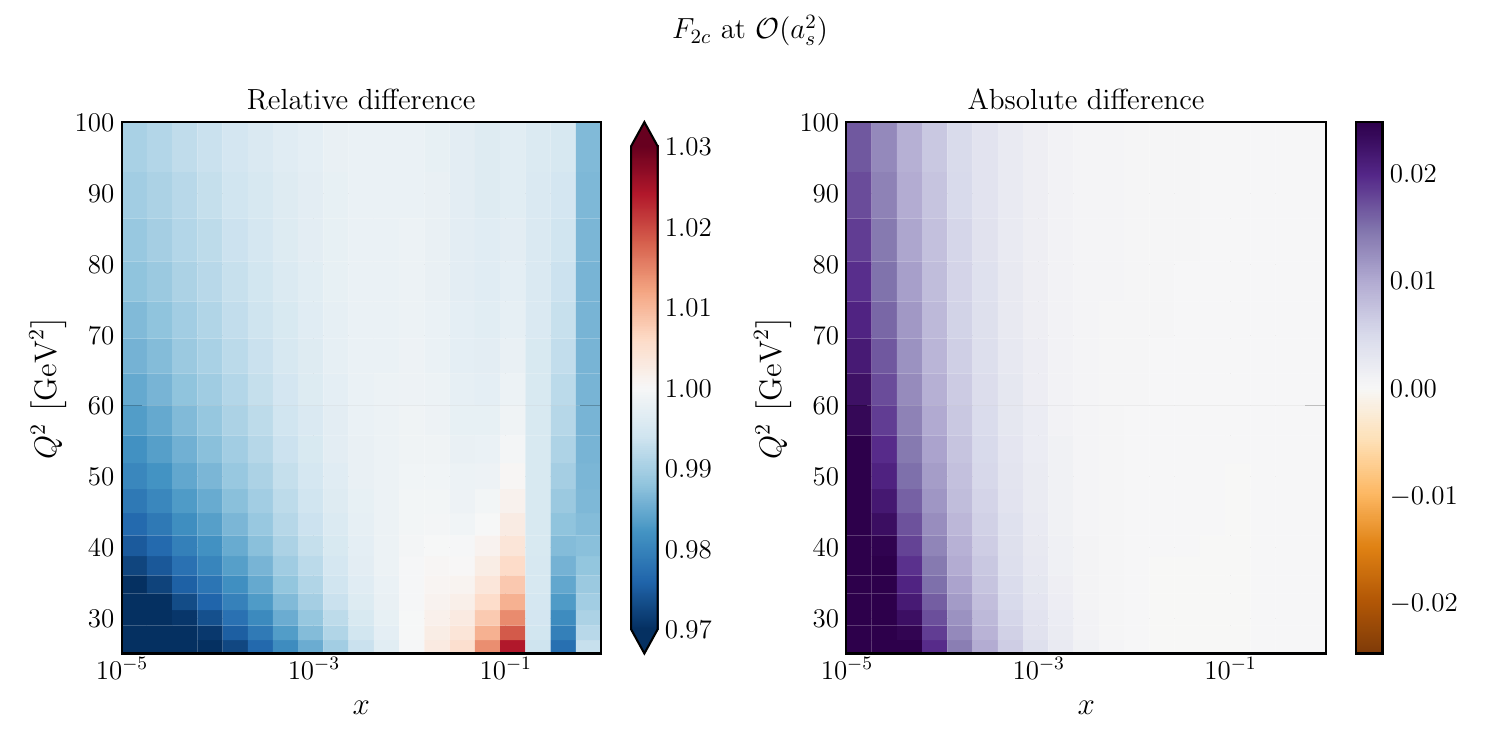}
    \caption{
        Relative (left) and absolute (right) difference of the charm-tagged structure function
        $F_{2c}$ evaluated with the method presented in this work and the prescription
        of Refs.~\cite{Forte:2010ta,Ball:2015dpa}.
    }
    \label{fig:F2_charm_bench}
\end{figure}

\begin{figure}
    \centering
    \includegraphics[width=\textwidth]{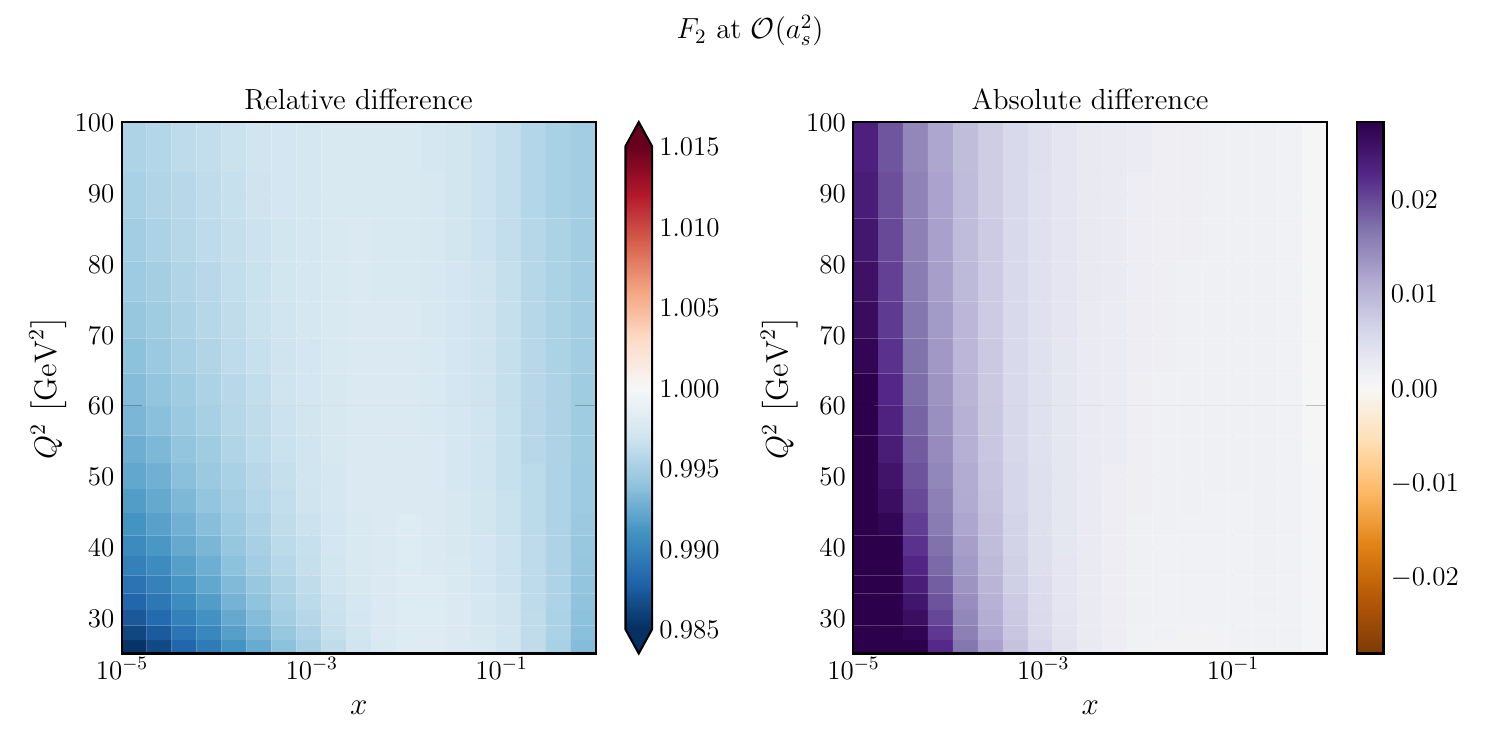}
    \caption{
        Same as \cref{fig:F2_charm_bench}, but now for the total structure function $F_2$.
    }
    \label{fig:F2_tot_bench}
\end{figure}

In \cref{fig:F2_charm_bench} we show the difference on the FONLL
charm tagged structure function $F_{2c}$, as a function of $x$ and $Q^2$,
evaluated with the new prescription described above and the one presented
in Refs.~\cite{Forte:2010ta,Ball:2015dpa}. We indicate both, the relative difference (left panel) and
absolute difference (right panel) in the considered kinematic range of $x \in [10^{-5},1]$ and
$Q^2 \in [25,100]~\si{\GeV^2}$.
The calculations are performed at NNLO accuracy and use as boundary condition \texttt{NNPDF40\_nnlo\_as\_01180} at $Q^2=1.65\ \mathrm{GeV}$ (and thus $n_f=4$).~\cite{NNPDF:2021njg}.
The same comparison for the total structure function $F_2$ is displayed in
\cref{fig:F2_tot_bench}.
While we observe a good overall agreement between the two prescriptions, in the small-$x$ region differences can be significant. This difference can be directly associated to the
collinear resummation: while the prescription of Refs.~\cite{Forte:2010ta,Ball:2015dpa}
relies on the finite order expansion of the resummation, entering through the inverse scheme change
in \cref{eq:Bdef}, the prescription of this paper retains the full resummation, by using the full
EKOs in \cref{eq:sF3,eq:sF4,eq:sF34}. These resummation terms are precisely the higher order
difference between the two prescriptions.
Performing the comparison at NLO accuracy or
with input PDFs containing a fitted charm component~\cite{Ball:2022qks} yields similar conclusions.

\subsection{Two masses case}
\label{sec:two}

\begin{figure}
  \centering
  \includegraphics[width=0.55\textwidth]{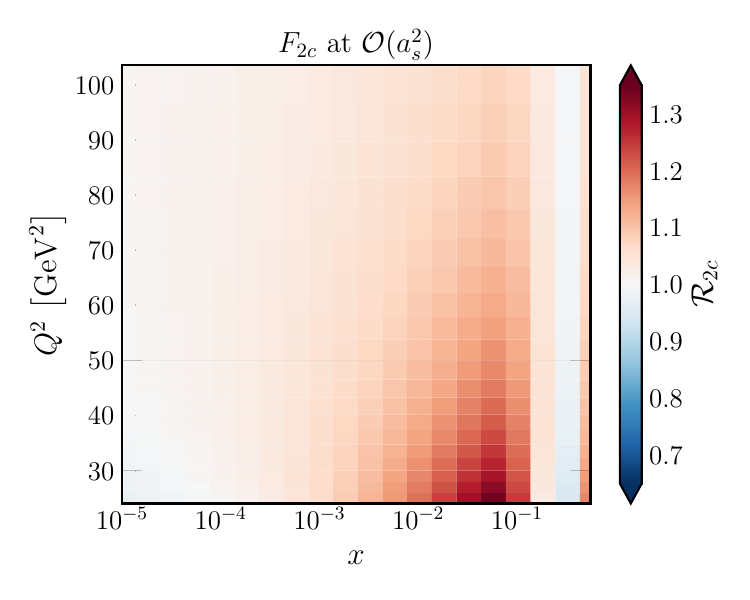}
  \caption{The ratio $\mathcal{R}_{2c}$ of \cref{eq:rc} representing the relative impact of the heavy charm contribution to $F_{2c}^{\mathrm{FONLL}}$
  at scales above the bottom quark mass.}
  \label{fig:rc}
\end{figure}

For the case of multiple masses it is possible to apply the prescription of Ref.~\cite{Forte:2010ta}
consecutively for each quark mass. However, this only yields faithful results if the quark masses are
strongly ordered (e.g.\ $m_c \ll m_b$), which is not realized in nature.
In particular, the charm mass corrections remains quite large for scales above the bottom
mass. This is shown explicitly in \cref{fig:rc} in which we plot the ratio
\begin{equation}
  \mathcal{R}_{2c} \equiv \frac{F_{2c}^{(4)}(Q^2,m_c^2)}{F_{2c}^{\mathrm{FONLL}}(Q^2,m_c^2)},
  \label{eq:rc}
\end{equation}
where $F_{2c}^{\mathrm{FONLL}}$ is defined as in \cref{eq:FONLL1} for the structure function $F_{2c}$.
In the calculation we use a charm quark mass $m_c=1.51\ \mathrm{GeV}$ and a bottom quark mass of $m_b=4.92\ \mathrm{GeV}$,
meaning that the plotted kinematic domain corresponds to scales $Q^2 > m_b^2$.
It is clear that even in this region, contributions due to the heavy charm mass effects
are up to 30\% and thus can not be neglected.

To produce accurate predictions, we therefore need to include simultaneously the massive effects of multiple heavy quarks. To account for both the charm and bottom masses, the approach of \cite{Forte:2010ta}
would require rewriting both FFNS3c and FFNS4b contributions in terms of FFNS5 (see \cref{eq:Bdef}), while
in our new method the inclusion is significantly simpler.

In practice we need to combine FFNS3c, where power-like terms for charm are
present, FFNS4b, where power-like terms for bottom are present, and FFNS5, where
collinear resummation for both quarks are fully accounted for. Using again set
theory language we write
\begin{align}
    F^{\mathrm{FONLL}}(Q^2,m_c^2,m_b^2)
        &= F^{(3c)} (Q^2, m_c^2) + F^{(4b)} (Q^2, m_c^2, m_b^2) + F^{(5)}(Q^2, m_c^2, m_b^2) \nonumber \\
        &\hspace{20pt} - F^{(3c\cap 4b)} (Q^2, m_c^2) - F^{(4b\cap 5)} (Q^2, m_c^2, m_b^2) - F^{(3c\cap 5)} (Q^2, m_c^2, m_b^2) \nonumber \\
        &\hspace{20pt} + F^{(3c\cap 4b\cap 5)} (Q^2, m_c^2, m_b^2)
        \label{eq:sFFONLL_2mass}
\end{align}
where the first line corresponds to the plain combination we want to combine,
the second line removes the double-counting between any two out of the three
schemes, and the third line adds back the triple-overlap between all schemes.
Note that here we are considering a consecutive approach, where we only consider
one massive quark at a time, however, we comment on the case of simultaneous
heavy quark effects in \cref{app:b}.

The expression for FFNS3c is given in \cref{eq:sF3}, while for FFNS4b we have
\begin{align}
    F^{(4b)} (Q^2, m_c^2, m_b^2) & = C^{(4b)}(Q^2/m_b^2) f^{(4b)}(Q^2) \\
    & = C^{(4b)}(Q^2/m_b^2) E^{(4b)}(\log(Q^2/m_c^2)) A^{(4b)}(0) f^{(3)}(m_c^2) \ ,
    \label{eq:mF4}
\end{align}
where $C^{(4b)}(Q^2/m_b^2)$ contains both logarithm and powers of $Q^2/m_b^2$,
but has no dependence the charm mass. For FFNS5 we have, similarly to \cref{eq:sF4},
\begin{align}
    F^{(5)} (Q^2, m_c^2, m_b^2) & = C^{(5)}f^{(5)}(Q^2) \\
    &= C^{(5)}E^{(4)}(\log(Q^2/m_b^2)) A^{(5)}(0) E^{(4)}(\log(m_b^2/m_c^2)) A^{(4)}(0) f^{(3)}(m_c^2)
    \label{eq:mF5}
\end{align}
where $C^{(5)}$ is a massless coefficient function and we suppress again all dependency
on the strong coupling $a_s$.

In analogy to the single mass case, the neighboring overlaps $F^{(3c \cap 4b)}$ and
$F^{(4b \cap 5)}$ are obtained by
\begin{align}
    F^{(3c \cap 4b)}(Q^2, m_c^2) &= C^{(3c \cap 4b)}(\log(Q^2/m_c^2)) f^{(3c)}(Q^2) \\
    &= C^{(3c \cap 4b)}(\log(Q^2/m_c^2)) E^{(3c)}(\log(Q^2/m_c^2)) f^{(3)}(m_c^2) \,,
\end{align}
and
\begin{align}
    F^{(4b \cap 5)}(Q^2, m_b^2) &= C^{(4b \cap 5)}(\log(Q^2/m_b^2)) f^{(4b)}(Q^2) \\
    & = C^{(4b \cap 5)}(\log(Q^2/m_b^2)) E^{(4b)}(\log(Q^2/m_c^2)) A^{(4b)}(0) f^{(3)}(m_c^2) \,,
\end{align}
where the corresponding coefficient functions are given by
\begin{align}
        C^{(3c \cap 4b)}(\log(Q^2/m_c^2)) &= C^{(4)} A^{(4)}(\log(Q^2/m_c^2)) = C^{(3c \cap 4)}(\log(Q^2/m_c^2)) \,,\\
        C^{(4b \cap 5)}(\log(Q^2/m_b^2)) &= C^{(5)} A^{(5)}(\log(Q^2/m_b^2)).
\end{align}

Before discussing the remaining double overlap $F^{(3c \cap 5)}$ in \cref{eq:sFFONLL_2mass}
we turn to the triple overlap $F^{(3c \cap 4b \cap 5)}$ for which we observe
the following identity:
\begin{equation}
\label{eq:trov}
    F^{(3c \cap 4b \cap 5)} = F^{(3c \cap 4b \cap 4b \cap 5)} = F^{((3c \cap 4) \cap (4b \cap 5))} 
\end{equation}
where for the first equality we used the fact that the intersection of a set with itself is an identity
operation in set theory and for the second equality we used the commutativity of the intersection
operation and the fact that $F^{(3c \cap 4b)} = F^{(3c \cap 4)}$, as
FFNS3c does not contain any bottom mass corrections as the bottom is decoupled.

From \cref{eq:trov} we recover the well known fact that the matching procedure is a sequential procedure,
i.e.\ we first add the charm contributions and then the bottom contributions.
This immediately implies that $F^{(3c \cap 5)} = F^{(3c \cap 4b \cap 5)}$ as the former would need to go through
the charm matching in any case. This reduces the number of contributions in \cref{eq:sFFONLL_2mass} in practice to just
the first five terms.

Nevertheless, using similar arguments as before we can give an explicit expression for the remaining double overlap
$F^{(3c \cap 5)}$ in \cref{eq:sFFONLL_2mass}
and write
\begin{align}
    F^{(3c \cap 5)} (Q^2, m_c^2, m_b^2) &= C^{(3c \cap 5)}(\log(Q^2/m_c^2),\log(Q^2/m_b^2)) f^{(3c)}(Q^2) \\
     &= C^{(3c \cap 5)}(\log(Q^2/m_c^2),\log(Q^2/m_b^2)) E^{(3c)}(\log(Q^2/m_c^2)) f^{(3)}(m_c^2) \,,
\end{align}
where the coefficient function
\begin{equation}
    C^{(3c \cap 5)}(\log(Q^2/m_c^2),\log(Q^2/m_b^2)) = C^{(5)} A^{(5)}(\log(Q^2/m_b^2)) A^{(4)}(\log(Q^2/m_c^2)) \, ,
\end{equation}
can again be found by shifting both the charm and bottom matching scale to $Q^2$.

Generalizing these expressions to the case of more than two masses is straightforward but tedious.
One would just follow the same procedure together with the corresponding set theory combination.
Moreover, the three mass case is less phenomenologically relevant than the two mass case:
the strange quark has a low enough mass to be always considered massless,
and the top quark mass is much larger than the bottom quark mass.

\subsection{A combined prescription for all scales}
\label{sec:damping}

We now address the issue of combining the above discussion into a combined prescription
which can be applied at all scales $Q^2$.

While one possible choice would be to use the prescription of \cref{sec:two} at all scales, this turns out to be
phenomenologically not well-behaved. Indeed, when $Q^2$ is smaller than one of the relevant quark masses,
the resummation of the corresponding collinear logarithms is power suppressed and the fixed order calculation
is reliable. In fact, also the prescription of Ref.~\cite{Forte:2010ta} is meant to be used only with $Q^2 > m^2_h$.
In our case this means that we use the single mass case described in \cref{sec:one} after crossing the first quark
threshold and the two mass case described in \cref{sec:two} after crossing the second quark threshold.
For scales higher than the top mass, one can, as mentioned before, apply an extension of the FONLL prescription for a
simultaneous treatment of charm, bottom, and top mass effects;
though because $m_c,m_b \ll m_t$, the effects of $m_c$ and $m_b$ are small at such high scales
and for simplicity, one may use a prescription neglecting charm and bottom mass effects.
Thus we have
\begin{equation}
\label{eq:finalcombination}
    F(Q^2, m_c^2, m_b^2) = \left\{ \begin{array}{ll}
        F^{(3cb)}(Q^2, m_c^2, m_b^2) &\text{if}~ Q^2 < m_c^2\\
        F^{(3c)}(Q^2, m_c^2) + F^{(4b)}(Q^2, m_c^2, m_b^2) - F^{(3c\cap 4)}(Q^2, m_c^2) &\text{if}~ m_c^2 \leq Q^2 < m_b^2\\
        \text{\cref{eq:sFFONLL_2mass}} &\text{if}~ m_b^2 \leq Q^2 < m_t^2 \\
        F^{(6)}(Q^2) &\text{if}~ Q^2 \geq m_t^2
    \end{array} \right.,
\end{equation}
where for $Q^2 < m_c^2$ charm mass corrections
are dealt in a FFNS setup (i.e.\ $F^{(3cb)}$) and above $m_c$ the FONLL prescription applies.
The same holds for the bottom mass effects: below $m_b$ bottom mass corrections are dealt in a FFNS
setup by either $F^{(3cb)}$ or $F^{(4b)}$ and above $m_b$ the FONLL prescription applies.

At low perturbative order it was found~\cite{Forte:2010ta} that in the single mass case the cancellation between the
FFNS4 structure functions, \cref{eq:sF4}, and the overlap, \cref{eq:sF34}, is incomplete near the heavy quark
threshold. To make the definition of the structure function continuous across the mass theresholds, Ref.~\cite{Forte:2010ta}
introduces a damping function
\begin{equation}
    \chi(Q^{2}, m^{2}) = \biggl(1 - \frac{m^{2}}{Q^{2}} \biggr)^{2} \,.
\end{equation}
For example, we can then modify \cref{eq:sFFONLL_2mass} to be
\begin{align}
    F^{\mathrm{FONLL}}(Q^2,m_c^2,m_b^2)
        &= F^{(3c)} (Q^2, m_c^2) + \chi(Q^{2}, m_{c}^{2})\biggl[ F^{(4b)} (Q^2, m_c^2, m_b^2) - F^{(3c\cap 4b)} (Q^2, m_c^2) \nonumber \\
        &\hspace{80pt}+ \chi(Q^{2}, m_{b}^{2}) \bigl[ F^{(5)}(Q^2, m_c^2, m_b^2) - F^{(4b\cap 5)} (Q^2, m_c^2, m_b^2)\bigr]\biggr]\,,
    \label{eq:sFFONLL_2mass_damp}
\end{align}
to ensure a smooth transition at the bottom threshold in \cref{eq:finalcombination}.

We briefly comment on the practical implementation of our new prescription in \cref{app:implementation}.

\section{Generalizations}
\label{sec:schemes}
In the following we discuss a series of generalizations which demonstrate the improved
applicability of our new prescription in comparison to Ref.~\cite{Forte:2010ta}.
We stress that all of these considerations are also possible in their prescription, but we would like to point out
the advantages gained on both theoretical grounds and practical grounds
when implementing the FONLL scheme using our new prescription.

\paragraph*{Perturbative and logarithmic accuracy}

It is the
limited perturbative knowledge of the various ingredients ($C,\beta,\gamma,A,d$) which make a
GM-VFNS prescription necessary. In fact, at all-order accuracy \cref{eq:sF4,eq:sF34}
would be identical and we would just be left with \cref{eq:sF3}.
However, in practice we have to work at a finite order and thus
a practical limitation of the prescription of Ref.~\cite{Forte:2010ta} is the required perturbative expansion of \cref{eq:Bdef},
which becomes technically challenging with increasing perturbative and/or logarithmic accuracy.

To give an example: the recently presented extraction of PDFs at aN$^3$LO~\cite{NNPDF:2024nan} requires a correct
treatment of DIS structure functions at the quoted accuracy. While it is possible~\cite{Zanoli} to use the
prescription of Ref.~\cite{Forte:2010ta}, the resulting formulas are lengthy and error prone. Instead, in our
new prescription no additional transformations are needed, as we obtained a clear separation between
evolution kernels and coefficient functions. The new prescription does not use \cref{eq:Bdef} but only relies
on the correct determination of the overlap coefficient function \cref{eq:C3c4}, which is needed in either prescription.

\paragraph*{Mixing accuracies}

As was pointed out in Ref.~\cite{Forte:2010ta} the naive first non-vanishing order for FFNS3c and FFNS4 differ,
and this is directly related to the different
FONLL schemes (FONLL-A, FONLL-B and FONLL-C) described in Ref.~\cite{Forte:2010ta}, since
they correspond to different accuracies of \cref{eq:sF3,eq:sF4} respectively.
Especially the case of FONLL-B requires additional attention as it relies on the observation that
the first non-vanishing order in the strong coupling is different for heavy quark production in massive and massless schemes when neglecting intrinsic heavy quark contributions.

For example, in DIS, the former emerges by photon-gluon fusion which opens at $O(a_s)$, the latter is
just ordinary photon-quark fusion that first contributes at $O(a_s^0)$.
Extending the computation to higher perturbative orders in our new prescription
is straightforward. Given the coefficients in FFNS3c, \cref{eq:sF3}, and
FFNS4, \cref{eq:sF4}, at a certain order, one just needs to identify the collinear
logarithmically order in FFNS3c such that only those also present in FFNS4 are
retained in the overlap~\cref{eq:sF34}.

\paragraph*{Changing the matching point}

In the above discussion we always assumed that the transition scale $\mu_h$ between the FFNS in which the
heavy quark $h$ is considered massive and the FFNS in which the same quark is considered massless
coincides with its mass $\mu_h=m_h$ -- this is a natural choice and indeed the most common convention.
However, this is still a choice and one may vary this scale to investigate its dependence.
In our new prescription this freedom is naturally embedded by the requirement to compute
ingredients which all follow a strict FFNS definition and thus include by definition this matching procedure.
Indeed, the full dependency is contained in the evolution operators $E$ as they naturally define the
transition between schemes.

\paragraph*{Treatment of intrinsic heavy quarks}

While Ref.~\cite{Forte:2010ta} only considers the case of perturbative generation of heavy quarks
its prescription was later extended to the presence of intrinsic heavy quark components in Ref.~\cite{Ball:2015dpa}.
In fact, the authors realized that this generalization simplifies the prescription of Ref.~\cite{Forte:2010ta}
because several intermediate terms now cancel amongst each other, as the first non-vanishing order is readjusted.

In our new prescription the inclusion of intrinsic heavy quark contributions fall naturally into place as they
simply contribute to the FFNS3c structure function, \cref{eq:sF3}, and hence also to the overlap, \cref{eq:sF34}.

\paragraph*{Observables beyond structure functions and with more collinear distributions}

Due to the clear separation between the partonic coefficient functions and the contributions from collinear evolution
our new prescription can be applied easily to any observable which may also contain more than one collinear
distribution (such as e.g.\ observables at the LHC~\cite{Cacciari:1998it}). We can give a straight generalization of \cref{eq:FONLL1} for
any observable $\sigma$:
\begin{equation}
    \sigma^\mathrm{FONLL}(Q^2,m_c^2) = \sigma^{(3c)}(Q^2,m_c^2) + \sigma^{(4)}(Q^2,m_c^2) - \sigma^{(3c\cap 4)}(Q^2,m_c^2)
\end{equation}
Just as before, $\sigma^{(3c)}$ and $\sigma^{(4)}$ have to be computed in a strict FFNS3c and FFNS4 setup respectively,
and $\sigma^{(3c\cap 4)}$ can be deduced from $\sigma^{(3c)}$ by identifying the collinear logarithms which are
fully resummed in the FFNS4 setup. If the observable $\sigma$ is not sensitive to heavy quark mass effects,
$\sigma^{(3c)}$ and $\sigma^{(3c\cap 4)}$ cancel each other exactly, as the former does not develop
any collinear logarithms.

\paragraph*{Longitudinally polarized structure functions}

Finally, a straight-forward generalization can be obtained for the case
of longitudinally polarized structure functions~\cite{Hekhorn:2024tqm} as, in fact, all
of the discussion also holds for a given spin state.

\section{Conclusion}
\label{sec:summary}

We have presented a new prescription of the FONLL scheme, which is perturbatively equivalent to the former prescription~\cite{Forte:2010ta} but
can be implemented more efficiently. The new prescription relies on the existence of coexisting flavor number PDFs,
i.e.\ an explicit decoupling of the number of active quarks and the factorization scale.

These PDFs can in practice be accessed using the \texttt{EKO} program in an efficient
way, while the coefficient functions can be computed using
\texttt{yadism}~\cite{Candido:2024rkr}.
We have implemented two ways how these numerical tools can be used to calculate predictions
such as the ones presented in the figures of this paper.
In particular, one can store the computed coefficients and
DGLAP kernels in PDF-independent FastKernel (FK) tables~\cite{Ball:2008by}
such that the computation can efficiently be performed for many different PDFs
without the need to repeat the often costly computation.

The new prescription features a clear separation between partonic matrix elements and evolution ingredients which
make it an ideal tool to investigate higher order corrections.
Moreover, our new prescription features a clear extension to the multi-mass case which is relevant, e.g., for
mid- and low-scale structure functions in DIS, such as the ones used in all modern PDF determinations.

The new prescription has been adopted by the NNPDF collaboration where it is used to compute predictions
for DIS observables~\cite{NNPDF:2024djq,NNPDF:2024dpb,NNPDF:2024nan,Hekhorn:2024tqm,Hekhorn:2024jrj,nnpdfpol20}.
Our new prescription does not imply further restrictions on the kinematically applicable domain
and, indeed, \cref{eq:finalcombination} gives a prescription which can be evaluated at all perturbative
scales. The actual kinematic cuts used in NNPDF fits are discussed in the respective references and
may differ for unpolarized and longitudinally polarized PDF extractions.

\section*{Acknowledgments}
We thank the members of the NNPDF collaboration for useful and insightful
discussions at the various stages of this project. We are grateful to Richard Ball and Stefano Forte for comments on the manuscript.

F.~H. is supported by the Academy of Finland project 358090 and is funded as a
part of the Center of Excellence in Quark Matter of the Academy of Finland, project 346326.
G.~M. is supported by NWO, the Dutch Research Council.
R.~S. is supported by the U.K. Science and Technology Facility Council (STFC)
consolidated grants ST/T000600/1 and ST/X000494/1.

\appendix
\section{Simultaneous two mass case}
\label{app:b}

In \cref{sec:two} we consider the case of a consecutive decoupling for charm and bottom,
i.e.\ we add heavy quark effects one at a time while still accounting for the other.
However, the FONLL scheme and, specifically, the new prescription we propose in this paper
can also be applied to the case where one considers simultaneous two mass
effects~\cite{Ablinger:2017err,Blumlein:2018jfm}.
In this case the massive corrections for either charm or bottom are always contributed
by the FFNS3cb structure function, while the collinear resummed parts do not carry any
mass dependency.

Eventually, we can give an analogue expression to \cref{eq:finalcombination}:
\begin{equation}
    F(Q^2, m_c^2, m_b^2) = \left\{ \begin{array}{ll}
        F^{(3cb)}(Q^2, m_c^2, m_b^2) &\text{if}~ Q^2 < m_c^2\\
        F^{(3cb)}(Q^2, m_c^2, m_b^2) + F^{(4)}(Q^2) - F^{(3cb\cap 4)}(Q^2, m_c^2, m_b^2) &\text{if}~ m_c^2 \leq Q^2 < m_b^2\\
        F^{(3cb)}(Q^2, m_c^2, m_b^2) + F^{(5)}(Q^2) - F^{(3cb\cap 5)}(Q^2, m_c^2, m_b^2) &\text{if}~ m_b^2 \leq Q^2 < m_t^2 \\
        F^{(6)}(Q^2) &\text{if}~ Q^2 \geq m_t^2
    \end{array} \right.,
\end{equation}
The overlap subtraction terms $F^{(3cb\cap 4)}$ and $F^{(3cb\cap 5)}$ follow from the same guiding
principle as the single mass case: they can only consist of collinear contributions of either just the charm quark (in the first case)
or both charm and bottom quarks (in the second case). Again, recall that the matching is done in a consecutive
manner and, thus, $F^{(5)}$ resums both charm and bottom collinear logarithms.

The implementation in the \texttt{yadism} library discussed in \cref{app:implementation} follows the prescription of \cref{sec:two}.

\section{Implementation}
\label{app:implementation}

The actual implementation of our new prescription requires the possibility
to compute coexisting flavor number PDFs, which is provided by the \texttt{EKO} library~\cite{Candido:2022tld},
and the computation of the partonic matrix elements using different
FFNS settings. In practice we apply our new prescription only to DIS
structure functions and we use the \texttt{yadism} library~\cite{Candido:2024rkr}
to provide the respective coefficient functions.

An explicit example of a DIS observable calculation using a fixed input PDF is
available in the \texttt{yadism} documentation at

\begin{center}
    {\bf \url{https://yadism.readthedocs.io/en/latest/overview/tutorials/fonll.html}~}.
\end{center}

Instead, if one may wish to compute the same set of observables for many different PDF sets,
renormalization or factorization scales, or values of
$\alpha_s$, as is done in PDF determinations or systematic parameter studies,
it is recommended to create interpolation grids in the
\texttt{PineAPPL} format~\cite{Carrazza:2020gss} that store the combination of
DIS coefficients and DGLAP evolution in PDF independent grids called fast-kernel
(FK) tables~\cite{Ball:2008by} which can be used to compute observables:
\begin{equation}
  \sigma = \sum_i \sum_a f_a(x_i,\mu_0^2)\mathrm{FK}_a(x_i,\mu_0^2).
\end{equation}

In this case the implementation is a multi-step process which is driven by
the \texttt{pineko} code, as illustrated at

\begin{center}
    {\bf \url{https://pineko.readthedocs.io/en/latest/theory/fonll.html}~}.
\end{center}

\bibliography{num_fonll}

\end{document}